# Improved Model for Wire-Length Estimation in Stochastic Wiring Distribution

Mohamed S. Hefeida
Electrical and Computer Engineering
University of Illinois at Chicago, Chicago, IL 60607, USA
Email: mhefei2@uic.edu

Masud H. Chowdhury
Computer Science and Electrical Engineering
University of Missouri – Kansas City, MO 64110
Email: masud@ieee.org

**Abstract:** This paper presents a pair of improved stochastic wiring distribution models for better estimation of on-chip wire-lengths. The proposed models provide 28% - 50% reduction in error when estimating the average on-chip wire length compared to the estimation using existing models. The impact of Rent's exponent on the average wire length estimation is also investigated to demonstrate limitations of the approximations used in some of the current models. To improve the approximations of the model a new threshold for Rent's constant is recommended. Simulation results demonstrate that proposed models with the new threshold reduce the error of estimation by 38% - 75% compared to the previous works.

## I. INTRODUCTION

THE dominance of interconnect parameters on the performance of deep submicron and nanoscale integrated circuits and systems is becoming ever more crucial with technology scaling [1]-[5]. For example, in 65nm technology circuit block performance shows 47% sensitivity to transistor parameters and 53% sensitivity to interconnect parameters [6]. In 130nm microprocessor, interconnects contribute over 50% of the dynamic power consumption, and repeaters consume almost half of the leakage power [7]. These performance figures are going to be further skewed towards interconnects as their density and number of layers increases (500nm technology used only 4 layers while 65nm technology uses 10 layers of interconnects [7]). Therefore, it is imperative that design and analysis of interconnects must be as accurate as possible.

Wire pitches for new high performance IC technologies are usually chosen using a stochastic wiring distribution that considers previous generations of technology and estimates the wire lengths of the chip to be designed with the logic technology. After the wire distribution is known, algorithms are used to find pitches of different interconnect levels based on limitations regarding cost and performance requirements. Some algorithms to predict die area and wire pitches that were used in older technologies can be found in [8]-[9]. Most of the current stochastic wire-length distributions show significant error when compared to actual data; hence, more accurate wire length estimates are necessary [7].

In this paper a set of improved model for stochastic wiring distribution is presented. The preliminary observations and analysis are presented in our conference submission [11]. Section II introduces some of the most popular existing stochastic wiring distribution models on which the proposed improvements are based. Some limitations and modifications of the existing models are revealed in the discussion of section II. Section III presents the new improved models along with the explanations of the modifications on the existing models to improve the average error in the interconnect wire length estimation. The effect of Rent's exponent on the average wire length estimation is also studied in this section. Finally, section IV concludes the paper.

## II. STOCHASTIC WIRING DISTRIBUTIONS

### A. Existing Wire Length Estimation Model

A number of publications have discussed stochastic wiring distributions. One of the most recent models is the one proposed in [7], which is based on the work in [12] with the assumption that gates in circuit block are randomly arranged rather than uniform distribution. A new quantity called a *gate socket* – a spot where a gate can be placed, was introduced. Any chip is considered to have many gate sockets, some of which are occupied by logic gates, the number of logic gates $N_{gates}$ is related to the number of sockets $N_{soc}$ as in (1) [7], where $P_{gates}$ is the percentage of the die area that is occupied by logic gates. The expected number of interconnects - $i(l)$ of a certain length $l$ is given as a product of $M(l)$ - the number of gate socket pairs separated by a distance $l$, and $I_{\exp}(l)$ - the average number of interconnects between a gate socket pair separated by $l$ as shown in (2). Following the derivation of the number of gate socket pairs separated by $l$, we obtain the expression of $M(l)$ as in (3) [7].

$$N_{gates} = N_{soc} \cdot P_{gates} \quad (1)$$

$$i(l) = M(l) \cdot I_{\exp}(l) \quad (2)$$

$$M(l) = \begin{cases} \dfrac{l^3}{3} - 2l^2\sqrt{N_{soc}} + 2\,l\,N_{soc} & 1 \le l \le \sqrt{N_{soc}} \\ \\ \dfrac{1}{3}\left(2\sqrt{N_{soc}} - l\right)^3 & \sqrt{N_{soc}} \le l \le 2\sqrt{N_{soc}} \end{cases} \quad (3)$$

$M(l)$ describes the total number of gate socket pairs separated by a distance $l$, assuming a $\sqrt{N_{soc}}$ by $\sqrt{N_{soc}}$ square array of gate sockets [7], [12]. The value of $l$ can be either expressed in terms of gate socket lengths or gate pitches. The gate socket length is the distance between two adjacent gate sockets and is equal to $((Die\ area)/N_{sockets})^{0.5}$ while the gate pitch is $((Die\ area)/N_{gates})^{0.5}$. Therefore, a gate socket length is equal to $(P_{gates})^{0.5}$ pitches [7].

The average number of interconnect between a gate socket pair separated by a distance $l$ is given by (4).

$$I_{exp}(l) = P(gate\ in\ block\ A) \cdot \frac{I_{A\text{-}to\text{-}C}}{N_C}, \quad (4)$$

where $P(gate\ in\ block\ A)$ is the probability that block A of Fig. 1 is occupied by a gate, $I_{A\text{-}to\text{-}C}$ is the average number of interconnects connecting block A to block C, and $N_C$ is the number of gates in block C as shown in (5) and (6) [7], [12].

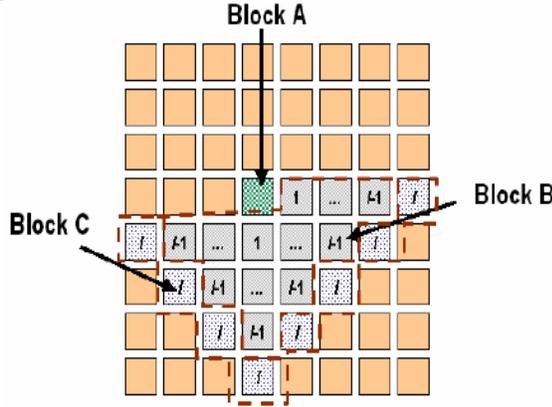

Fig. 1. Block definitions for exact wire length distribution [7]

$$P(gate\ in\ block\ A) = \frac{N_{gates}}{N_{soc}} = P_{gates} \quad (5)$$

$$I_{A\text{-}to\text{-}C} = \alpha k \left[ (N_A + N_B)^p - N_B^p + (N_B + N_C)^p - (N_A + N_B + N_C)^p \right] \quad (6)$$

Here $\alpha = f.o/(f.o+1)$, $f.o$ is the average fan-out of the system, $k$ and $p$ are Rent's constant and Rent's exponent respectively, and $N_A$ and $N_B$ are the number of gates in blocks A and B respectively. Considering gates to be randomly distributed in gate sockets, the following approximations hold since the number of gates in a block is simply related to the number of sockets through the probability $P_{gates}$. Therefore, $N_A$, $N_B$ and $N_C$ can be found as in (7).

$$\begin{aligned} N_A &= 1 \\ N_B &= P_{gates}(l^2 - l) \\ N_C &= 2lP_{gates} \end{aligned} \quad (7)$$

The average number of interconnects of length $l$ (in gate socket lengths) can be obtained by combining (2) through (7), which leads to the expressions shown in (8), (9), and (10). Here $i(l)$ shown in (8) gives the average number of interconnects between a gate socket pair separated by a distance $l$ for different range of number of sockets ($N_{soc}$). The normalization factor $\Gamma$ is introduced to compensate for the partial Manhattan circle approximation used in the distribution [12].

$$i(l) = \begin{cases} \dfrac{\alpha k \Gamma}{2} \left( \dfrac{l^3}{3} - 2l^2\sqrt{N_{soc}} + 2lN_{soc} \right) l^{2p-4} & 1 \leq l < \sqrt{N_{soc}} \\ \\ \dfrac{\alpha k \Gamma}{6} \left( 2\sqrt{N_{soc}} - l \right)^3 l^{2p-4} & \sqrt{N_{soc}} \leq l < 2\sqrt{N_{soc}} \end{cases} \quad (8)$$

$$\Gamma = \frac{2N_{gates}(1 - N_{gates}^{p-1})}{-N_{soc}^p \dfrac{1+2p-2^{2p-1}}{p(p-1)(2p-1)(2p-3)} - \dfrac{1}{6p} + \dfrac{2\sqrt{N_{soc}}}{2p-1} - \dfrac{N_{soc}}{p-1}} \quad (9)$$

$$L_{avg}(in\ gate\ socket\ lengths) = \frac{\int_1^{2\sqrt{N_{soc}}} l\, i(l)\, dl}{\int_1^{2\sqrt{N_{soc}}} i(l)\, dl} \quad (10)$$

### B. Limitation and Modification of Existing Model

The model presented in [7] is an improved version of the model presented in [12]. Although model of [7] provides reduction in estimation error compared to the original model of [12], the presented results are optimistic and the error calculation is biased towards reducing it, (i.e. if the error is calculated accurately it will result in more modest results). For better estimation of wire length a more accurate calculation of the error in estimating $L_{avg}$ is required. With this motivation we have solved the integral in (10) to obtain a model for $L_{avg}$. The resulting model of this solution is as shown in (11). The detailed steps of this solution are shown in *APPENDIX A* at the end of this paper. Based on this derivation we present an improved and corrected version of the model of [7] as in (11). It can be observed that the modified model in (11) is similar to that derived in [7] multiplied by $\dfrac{\sqrt{N_{soc}}}{(p-0.5)}$. Interestingly, we have observed that the optimistic results presented in [7] can not be obtained by directly using their model. However, the results presented in [7] can actually be obtained by the modified of (11). If the probability $P_{gates}$ is equal to 1 (i.e. $N_{soc}=N_{gates}$), the results match those presented in [12]. It is also useful to note that the results obtained using (11) will estimate the average wire length in gate socket lengths, which should be related to the gate pitches by a factor of $\sqrt{P_{gates}}$, i.e. to get $L_{avg}$ in terms of gate pitches we simply multiply the results obtained in terms of gate socket lengths by $\sqrt{P_{gates}}$.

$$L_{avg} = \frac{\sqrt{N_{soc}}}{(p-0.5)} \frac{\left[\frac{p-0.5}{p} - \sqrt{N_{soc}} - \frac{p-0.5}{6(p+0.5)\sqrt{N_{soc}}} + N_{soc}^p \left(\frac{-p-1+4^{p-0.5}}{2p(p+0.5)(p-1)}\right)\right]}{-N_{soc}^p \frac{1+2p-2^{2p-1}}{p(p-1)(2p-1)(2p-3)} - \frac{1}{6p} + \frac{2\sqrt{N_{soc}}}{2p-1} - \frac{N_{soc}}{p-1}} \quad (11)$$

$$L_{avg} = \frac{\sqrt{N_{soc}}}{(p-0.5)} \frac{\left[\frac{p-0.5}{p} - \sqrt{N_{soc}} - \frac{p-0.5}{6(p+0.5)\sqrt{N_{soc}}} + N_{soc}^p \left(\frac{-p-1+4^{p-0.5}}{2p(p+0.5)(p-1)}\right)\right]}{\left(-N_{soc}^p \frac{1+2p-2^{2p-1}}{p(p-1)(2p-1)(2p-3)} - \frac{1}{6p} + \frac{2\sqrt{N_{soc}}}{2p-1} - \frac{N_{soc}}{p-1}\right)(2P_{gates})^{0.25}} \quad (12)$$

$$L_{avg} = \frac{\sqrt{N_{gates}}}{(p-0.5)} \frac{\left[\frac{p-0.5}{p} - \sqrt{N_{gates}} - \frac{p-0.5}{6(p+0.5)\sqrt{N_{gates}}} + N_{gates}^p \left(\frac{-p-1+4^{p-0.5}}{2p(p+0.5)(p-1)}\right)\right]}{\left(-N_{gates}^p \frac{1+2p-2^{2p-1}}{p(p-1)(2p-1)(2p-3)} - \frac{1}{6p} + \frac{2\sqrt{N_{gates}}}{2p-1} - \frac{N_{gates}}{p-1}\right)(2P_{gates})^{0.25}} \quad (13)$$

TABLE I
ESTIMATING AVERAGE INTERCONNECT WIRE LENGTH, $L_{avg}$ USING OLD AND IMPROVED MODELS

| C-1 | C-2 | C-3 | C-4 | C-5 | C-6 | C-7 | C-8 | C-9 | C-10 | C-11 |
|---|---|---|---|---|---|---|---|---|---|---|
| # of gates | Rent's exponent (p) | Actual data | Davis [12] avg. length | %error | Modified Davis | %error | Sekar [7] $P_{gates}=0.75$ | %error | Modified Sekar $P_{gates}=0.75$ | %error |
| 2146 | 0.75 | 3.53 | 5.26 | 49.0085 | 4.423142 | 25.30147 | 4.8713 | 37.99717 | 4.4017 | 24.69405 |
| 576 | 0.75 | 2.98 | 3.9 | 30.87248 | 3.279516 | 10.05087 | 3.6008 | 20.83221 | 3.2537 | 9.184564 |
| 528 | 0.59 | 2.20 | 3.12 | 41.81818 | 2.623613 | 19.25514 | 2.7982 | 27.19091 | 2.5285 | 14.93182 |
| 671 | 0.57 | 2.63 | 3.12 | 18.63118 | 2.623613 | 0.24285 | 2.8251 | 7.418251 | 2.5528 | 2.935361 |
| 1239 | 0.47 | 2.14 | 2.96 | 38.31776 | 2.489068 | 16.31159 | 2.6474 | 23.71028 | 2.3922 | 11.78505 |
| 73 | 0.667 | 2.00 | 2.35 | 17.5 | 1.976118 | 1.1941 | 2.1437 | 7.185 | 1.9370 | 3.15 |
| 78 | 0.667 | 2.27 | 2.38 | 4.845815 | 2.001345 | 11.835 | 2.1704 | 4.387665 | 1.9612 | 13.60352 |
| 72 | 0.667 | 1.88 | 2.34 | 24.46809 | 1.967709 | 4.665372 | 2.1381 | 13.72872 | 1.9320 | 2.765957 |
| 252 | 0.667 | 2.73 | 2.96 | 8.424908 | 2.489068 | 8.82535 | 2.7090 | 0.769231 | 2.4479 | 10.33333 |
| 236 | 0.667 | 2.198 | 2.93 | 33.303 | 2.463841 | 12.09468 | 2.6755 | 21.72429 | 2.4176 | 9.990901 |
| 237 | 0.667 | 2.887 | 2.93 | 1.489435 | 2.463841 | 14.6574 | 2.6776 | 7.253204 | 2.4195 | 16.19328 |
| 55 | 0.667 | 1.579 | 2.23 | 41.22863 | 1.87521 | 18.75934 | 2.0332 | 28.76504 | 1.8372 | 16.35212 |
| **59** | **0.667** | **1.38** | **2.25** | **63.04348** | **1.892028** | **37.10348** | **2.0600** | **49.27536** | **1.8614** | **34.88406** |
| 62 | 0.667 | 2.08 | 2.28 | 9.615385 | 1.917255 | 7.82428 | 2.0791 | 0.043269 | 1.8787 | 9.677885 |
| Avg error | | | | **27.3262** | | **13.86897** | | **17.87719** | | **12.89156** |

TABLE II
VALIDATING NEW MODIFIED MODELS

| $N_{gates}$ | Rent's expon. (p) | Actual data | Davis | Modified Davis | Modified Sekar $P_{gates}=0.5$ | Sekar $P_{gates}=0.75$ | Modified Sekar $P_{gates}=0.75$ |
|---|---|---|---|---|---|---|---|
| 55 | 0.583 | 1.579 | 2.119 | 1.7818 | 1.6784 | 1.9237 | 1.7383 |
| **59** | **0.502** | **1.384** | **2.0467** | **1.7211** | **1.5981** | **1.8475** | **1.6694** |
| 62 | 0.745 | 2.077 | 2.3926 | 2.0119 | 1.957 | 2.2001 | 1.988 |
| 72 | 0.648 | 1.877 | 2.3103 | 1.9427 | 1.853 | 2.1079 | 1.9047 |
| 78 | 0.768 | 2.27 | 2.552 | 2.1459 | 2.1006 | 2.3527 | 2.1259 |
| 236 | 0.581 | 2.198 | 2.6819 | 2.2552 | 2.1187 | 2.4324 | 2.1979 |
| 237 | 0.762 | 2.887 | 3.2412 | 2.7255 | 2.6809 | 2.9944 | 2.7058 |
| 252 | 0.713 | 2.736 | 3.1124 | 2.617 | 2.5414 | 2.8606 | 2.5849 |
| 1118 | 0.69 | 3.6 | 4.0879 | 3.4375 | 3.3243 | 3.7515 | 3.3898 |
| Avg. error | | | 21.59764 | 6.158651 | 6.866796 | 11.14872 | 6.717929 |

| TABLE III |||||||||
|---|---|---|---|---|---|---|---|---|
| ESTIMATING $L_{avg}$ USING APPROXIMATE MODELS PRESENTED IN [7] |||||||||
| # of gates | Rent's exponent | Actual data | Davis Approx. $L_{avg}$ | %error | Sekar Approx. $P_{gates}=0.5$ | %error | Sekar Approx. $P_{gates}=0.75$ | %error |
| 2146 | 0.75 | 3.53 | 4.8756 | 38.11 | 4.0999 | 16.14 | 4.5373 | 28.53 |
| 576 | 0.75 | 2.98 | 3.5094 | 17.76 | 2.9510 | 0.97 | 3.2658 | 9.590 |
| 528 | 0.59 | 2.20 | 6.9424 | 215.56 | 5.225 | 137.5 | 6.17 | 180.45 |
| 671 | 0.57 | 2.63 | 8.6461 | 228.74 | 6.4176 | 144.01 | 7.6400 | 190.49 |
| 1239 | 0.47 | NA | NA | NA | NA | NA | NA | NA |
| 73 | 0.667 | 2.00 | 3.2010 | 60.05 | 2.5412 | 27.06 | 2.9086 | 45.43 |
| 78 | 0.667 | 2.27 | 3.2366 | 42.58 | 2.5695 | 13.19 | 2.9410 | 29.55 |
| 72 | 0.667 | 1.88 | 3.1937 | 69.87 | 2.5354 | 34.86 | 2.9019 | 54.35 |
| 252 | 0.667 | 2.73 | 3.9368 | 44.20 | 3.1254 | 14.48 | 3.5772 | 31.03 |
| 236 | 0.667 | 2.198 | 3.8939 | 77.15 | 3.0913 | 40.64 | 3.5382 | 60.97 |
| 237 | 0.667 | 2.887 | 2.93 | 1.489 | 3.3399 | 15.68 | 3.0918 | 7.09 |
| 55 | 0.667 | 1.579 | 3.0532 | 93.36 | 2.4239 | 53.50 | 2.7743 | 75.69 |
| 59 | 0.667 | 1.38 | 3.0892 | 123.85 | 2.4525 | 77.71 | 2.8070 | 103.40 |
| 62 | 0.667 | 2.08 | 3.1149 | 49.75 | 2.4729 | 18.88 | 2.8303 | 36.07 |
| Avg error | | | | 75.89 | | 42.34 | | 60.9 |

## III. IMPROVED MATHEMATICAL MODELS

In addition to the proposed modification of the model of [12], the authors of [7] also proposed a set of simplified models with certain limitations on the gate count and Rent's exponent. For clarity we differentiate between these models by naming the exact models and the approximate models. In this section we attempt to minimize the estimation errors of both the exact and the approximate models.

### A. Improvement of Exact Models

Our modified model of (11) identifies the limitation of the existing model proposed in [7]. However, our analysis reveals that further improvement of these models can be achieved. Based on our analytical derivation and experimental observation we present a pair of new improved models as shown in (12) and (13). Close observation reveals that the new models of (12) and (13) can actually be obtained by dividing (11) by $(2P_{gates})^{0.25}$. Due to the limitation of space we have omitted the detail mathematical derivations of these models. The derivational steps are similar to those in Appendix-I.

To verify the models of (11), (12) and (13) we have used exactly the same set of data (columns C-1, C-2 and C-3 of Table-I) used in [7] for fair comparison. Using (11), which is a modified version of the model in [7], first we have calculated the average interconnect wire lengths ($L_{avg}$) (see column C-8 of Table-I). As mentioned earlier the results obtained by (11) are in terms of gate socket lengths. By multiplying (11) with the factor $\sqrt{P_{gates}}$ we obtained $L_{avg}$ in terms of gate pitches. The calculated results will be similar to the results in [12] when $P_{gates}=1$ (see column C-4 of Table-I). It is also worth mentioning that the error values (column C-5 and C-9) calculated here are more accurate than those presented in [7] as we consider error to be any deviation from the actual data. Here we are calculating the mean absolute error (MAE) rather than the average error, which can be misguiding when assessing our wire-length estimation model. Even if we calculate the average error, our model is more accurate than previous models. [7], [12].

Next, we have used the improved models (12) and (13) to calculate average wire length (see columns C-6 and C-10). From the comparison of the data it is observed that the new modified models – (12) and (13) improve the results and reduced the error significantly (see columns C-7 and C-11). The error is reduced by almost 50% in Davis's [12] distribution and in Sekar's [7] by 28% when $P_{gates}=0.75$.

The comparison in Table-1 is based on the exact same data of [7]. To further validate the proposed improved models a set of practical data from [13] and [14] is used to estimate average interconnect wire lengths (in gate pitches) using the old and the proposed models. The actual data from [13] and [14] and the results are presented in Table-II, which shows that the modified exact model decreased the error by approximately 75% and 39% from when applied to models proposed in [12] and [7] respectively. It should be noted that the results obtained from [7] and the new modified version will be similar for $P_{gates}=0.5$. Comparison of the data in Table-I and Table-II shows that the value of $p$ used in [7] was purely based on approximation, and in the worst case it is assumed to be in the range from 0.502 to 0.667. This is a poor approximation that will lead to inaccurate results. The validity of this approximation is investigated in the next subsection.

### B. Limitations of the Approximate Models

We have also examined the approximate models of [7] shown in (14) and (15). A comparison of the average wire length estimated using these two approximate models and the values obtained from [7] are shown in Table-III. It can be observed that the approximation is poor, and gives significant error when compared to both actual data and data estimated using the exact models of [7]. This approximation may be useful in special cases, but not for general cases. It shows some individual improvements, but not consistent. For example, the approximated model shows a significant improvement in error when the number of gates is 576 and Rent's exponent $p=0.75$; while when the number of gates is increased to only 671 and Rent's $p$ decreased to 0.57 we observe a significant increase in error although the approximation is based on the condition that $p>0.5$ and the number of gates is large.

$$L_{avg} = N_{gates}^{p-0.5}\left(\frac{p+1-4^{p-0.5}}{2p(p-0.5)(p+0.5)}\right) \quad (14)$$

$$L_{avg} = P_{gates}^{1-p} N_{gates}^{p-0.5}\left(\frac{p+1-4^{p-0.5}}{2p(p-0.5)(p+0.5)}\right) \quad (15)$$

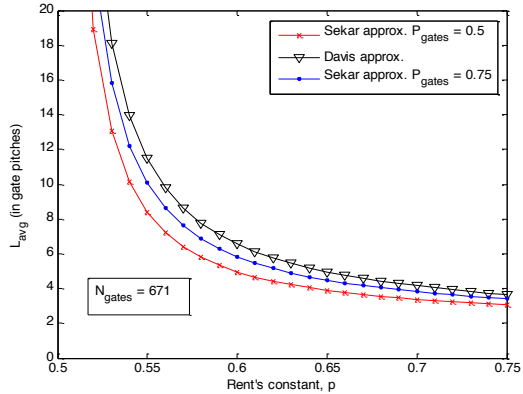

Fig. 2: Effect of Rent's exponent on the average interconnect-length for a large number of gates

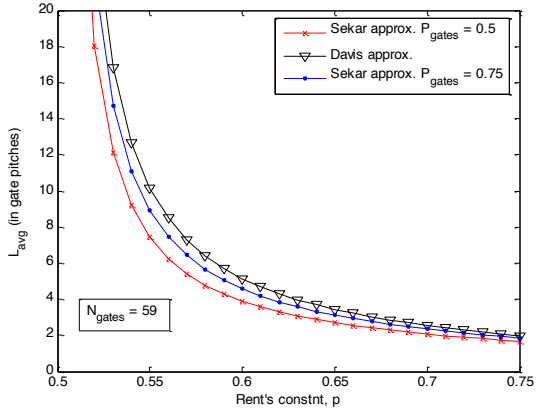

Fig. 3: Effect of Rent's exponent on the average interconnect length for a relatively small number of gates

From Table-III, it is clear that the approximation needs to be refined. To further emphasize the problem we plot $L_{avg}$ vs. $p$ to show its significant effect when using the approximated models, and suggest a new threshold for the value of $p$ based on the results shown in Fig. 2 and Fig. 3, which show a dramatic change in $L_{avg}$ when Rent's exponent ($p$) is in the range of 0.5 to 0.65, and above 0.65 the rate of change decreases significantly. Therefore, we suggest 0.65 as a new threshold value for $p$ above which the approximate models are more accurate.

### C. Analyzing the Impact of Number of Gates and Rents Constant

The number of gates in Fig. 2 and Fig. 3 were chosen according to the maximum error in large and small number of gates as shown in Table-III. The gate count does not make a dramatic difference in $L_{avg}$ as $p$ does. This is further illustrated in Fig. 4. Comparing the effects of the number of gates and Rent's exponent on $L_{avg}$ shows the different effect of each of them. From Fig. 4 it is observed that if the number of gates is increased 4x it makes at the most a 41.4% difference in the value of $L_{avg}$, while changing $p$ by less than 2% leads to a change of 49% in $L_{avg}$ as shown in Fig. 3.

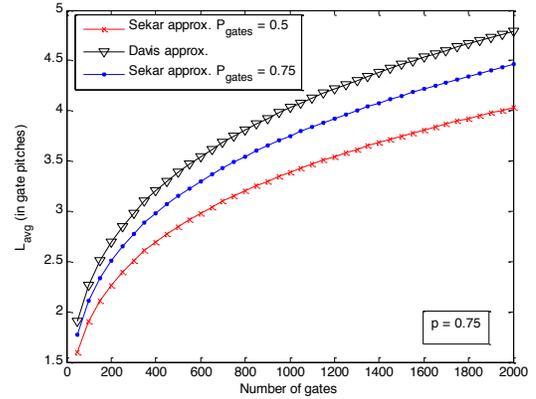

Fig. 4: Effect of gate count on the average interconnect wire length

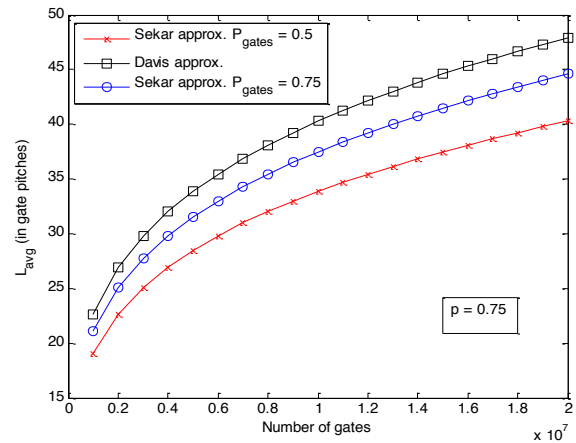

Fig. 5: Effect of gate count on the average interconnect length over an extended range

Today's single core microprocessor would likely have 20M gates and a logic die area of 20 mm$^2$ [15]. Future, multi-core system-on-chips (SOC) are expected to have nearly 100M gates in die area of 400mm$^2$. In these scenarios whether the existing wire length models would work or not is an issue to be investigated. Therefore, we extended the range of Fig. 4 to examine the behavior of the different models as shown below in Fig. 5. It is observed that with the increasing number of gate counts and wire lengths the estimation error using existing models increases.

### D. Modifying Approximate Models

In our analysis it is observed that if the same approximation as in the exact models is used, the approximated models provide only a 10% improvement in the error. Based on the analysis and observation we propose the following pair of modified approximate models (16) and (17), from the results of Table-IV it is observed that significant improvement in error can be obtained using them.

$$L_{avg} = N_{gates}^{p-0.5}\left(\frac{p+1-4^{p-0.5}}{2p(p-0.5)(p+0.5)}\right)\frac{1}{\sqrt{2P_{gates}}} \quad (16)$$

$$L_{avg} = P_{gates}^{1/2-p} N_{gates}^{p-0.5}\left(\frac{p+1-4^{p-0.5}}{2\sqrt{2}p(p-0.5)(p+0.5)}\right) \quad (17)$$

Application of both the new suggested threshold for Rent's exponent and the proposed modification on the approximated models shows significant reduction in error as shown in Table-V when compared to the results in Table III. The average error was reduced by almost a factor of 4 compared to Davis's models, while compared to Sekar's models a reduction in error by a factor of 3 is obtained.

TABLE IV
RESULTS OF MODIFIED APPROXIMATE MODELS

| # of gates | Rent's exponent | Actual data | Modified Approx. Davis'. | %error | Mod. approx. Sekars' $P_{gates}=.5$ | %error | Mod. approx. Sekars' $P_{gates}=0.75$ | %error |
|---|---|---|---|---|---|---|---|---|
| 2146 | 0.75 | 3.53 | 3.447603 | 2.334193 | 4.0999 | 16.14 | 3.70221 | 4.87847 |
| 576 | 0.75 | 2.98 | 2.481544 | 16.72671 | 2.9510 | 0.97 | 2.66385 | 10.6091 |
| 528 | 0.59 | 2.20 | 4.909065 | 123.139 | 5.225 | 137.5 | 5.02971 | 128.6232 |
| 671 | 0.57 | 2.63 | 6.113775 | 132.463 | 6.4176 | 144.01 | 6.2307 | 136.9087 |
| 1239 | 0.47 | NA | NA | NA | NA | NA | NA | NA |
| 73 | 0.667 | 2.00 | 2.263471 | 13.17355 | 2.5412 | 27.06 | 2.36586 | 18.293 |
| 78 | 0.667 | 2.27 | 2.288644 | 0.821322 | 2.5695 | 13.19 | 2.39295 | 5.4163 |
| 72 | 0.667 | 1.88 | 2.258309 | 20.12282 | 2.5354 | 34.86 | 2.36586 | 25.84362 |
| 252 | 0.667 | 2.73 | 2.783765 | 1.969414 | 3.1254 | 14.48 | 2.91669 | 6.838462 |
| 236 | 0.667 | 2.198 | 2.75343 | 25.26979 | 3.0913 | 40.64 | 2.88057 | 31.05414 |
| 237 | 0.667 | 2.887 | 2.071843 | 28.23543 | 3.3399 | 15.68 | 2.88057 | 0.22272 |
| 55 | 0.667 | 1.579 | 2.158959 | 36.72951 | 2.4239 | 53.50 | 2.2575 | 42.97023 |
| 59 | 0.667 | 1.38 | 2.184415 | 58.29094 | 2.4525 | 77.71 | 2.28459 | 65.55 |
| 62 | 0.667 | 2.08 | 2.202588 | 5.893654 | 2.4729 | 18.88 | 2.30265 | 10.70433 |
| Avg error | | | | 35.78228 | | 42.34 | | 37.53171 |

TABLE V
APPLYING THE NEW THRESHOLD AND THE NEW MODIFICATIONS ON THE APPROXIMATE MODELS

| # of gates | Rent's exponent | Actual data | Mod. Davis Approx. | %error | Mod. approx $P_{gates}=.5$ | %error | Mod. Approx. $P_{gates}=.75$ | %error |
|---|---|---|---|---|---|---|---|---|
| 2146 | 0.75 | 3.53 | 3.447603 | 2.334193 | 3.7 | 4.815864 | 3.70221 | 4.87847 |
| 576 | 0.75 | 2.98 | 2.481544 | 16.72671 | 2.66 | 10.73826 | 2.66385 | 10.6091 |
| 73 | 0.667 | 2.00 | 2.263471 | 13.17355 | 2.29 | 14.5 | 2.36586 | 18.293 |
| 78 | 0.667 | 2.27 | 2.288644 | 0.821322 | 2.32 | 2.202643 | 2.39295 | 5.4163 |
| 72 | 0.667 | 1.88 | 2.258309 | 20.12282 | 2.29 | 21.80851 | 2.36586 | 25.84362 |
| 252 | 0.667 | 2.73 | 2.783765 | 1.969414 | 2.82 | 3.296703 | 2.91669 | 6.838462 |
| 236 | 0.667 | 2.198 | 2.75343 | 25.26979 | 2.79 | 26.93358 | 2.88057 | 31.05414 |
| 237 | 0.667 | 2.887 | 2.071843 | 28.23543 | 2.79 | 3.359889 | 2.88057 | 0.22272 |
| 55 | 0.667 | 1.579 | 2.158959 | 36.72951 | 2.19 | 38.69538 | 2.2575 | 42.97023 |
| 59 | 0.667 | 1.38 | 2.184415 | 58.29094 | 2.21 | 60.14493 | 2.28459 | 65.55 |
| 62 | 0.667 | 2.08 | 2.202588 | 5.893654 | 2.23 | 7.211538 | 2.30265 | 10.70433 |
| Avg error | | | | 19.05158 | | 17.60975 | | 20.2164 |

## IV. CONCLUSION

This paper analyzes the limitations of existing models to estimate the average interconnect wire lengths in stochastic wiring distribution. Here a set of improved models is presented to improve accuracy of the wire length estimation. The new modified models provide 28% - 50% reduction in error when compared to the models presented in [7] and [12] respectively. It is illustrated that the approximation of Rent's exponent used in the existing works and its effect on the estimation of $L_{avg}$ have some limitations. Our analysis of the impact of Rent's constant on wire length estimation has led us to suggest a different threshold for Rent's exponent in order to be able to use the approximated models, and the error is reduced by 39% - 75% when using the new approximate models combined with the new threshold of Rent's constant for estimating the average wire length.

## APPENDIX

This appendix illustrates the analytical steps of estimating the average interconnect wire length $L_{avg}$.

$$i(l) = \begin{cases} \dfrac{\alpha k \Gamma}{2}\left(\dfrac{l^3}{3} - 2l^2\sqrt{N_{sockets}} + 2lN_{sockets}\right)l^{2p-4} & 1 \le l < \sqrt{N_{sockets}} \\ \dfrac{\alpha k \Gamma}{6}\left(2\sqrt{N_{sockets}} - l\right)^3 l^{2p-4} & \sqrt{N_{sockets}} \le l < 2\sqrt{N_{sockets}} \end{cases} \quad (A1)$$

$$L_{avg}\ (in\ gate\ socket\ lengths) = \frac{\int_{1}^{2\sqrt{N_{sockets}}} li(l)dl}{\int_{1}^{2\sqrt{N_{sockets}}} i(l)dl} \quad (A2)$$

We will first evaluate the numerator.

$$\int_{1}^{2\sqrt{N_{sockets}}} li(l)dl = \frac{\alpha k \Gamma}{2}\int_{1}^{\sqrt{N_{sockets}}}\left(\frac{l^3}{3} - 2l^2\sqrt{N_{sockets}} + 2lN_{sockets}\right)l^{2p-3}dl + \frac{\alpha k \Gamma}{6}\int_{\sqrt{N_{sockets}}}^{2\sqrt{N_{sockets}}}\left(2\sqrt{N_{sockets}} - l\right)^3 l^{2p-3}dl \quad (A3)$$

Let $P_1$ and $P_2$ be the first and second parts of the integral in (A3), therefore:

$$p_1 = \frac{\alpha k \Gamma}{2} \int_1^{\sqrt{N_{sockets}}} \left( \frac{l^{2p}}{3} - 2\sqrt{N_{sockets}} l^{2p-1} + 2N_{sockets} l^{2p-2} \right) dl \tag{A4}$$

$$p_1 = \frac{\alpha k \Gamma}{2} \left[ \frac{l^{2p+1}}{6p+3} - \frac{2\sqrt{N_{sockets}} l^{2p}}{2p} + \frac{2N_{sockets} l^{2p-1}}{2p-1} \right]_1^{\sqrt{N_{sockets}}} \tag{A5}$$

After substituting the integration limits in (A5) we get:

$$p_1 = \frac{\alpha k \Gamma}{6} \left( \frac{3N_{sockets}^{p+0.5}}{6p+3} - \frac{3N_{sockets}^{p+0.5}}{p} + \frac{3N_{sockets}^{p+0.5}}{p-0.5} - \frac{3}{6p+3} + \frac{3N_{sockets}^{0.5}}{p} - \frac{3N_{sockets}}{p-0.5} \right) \tag{A6}$$

Similarly we integrate $P_2$ to get:

$$p_2 = \frac{\alpha k \Gamma}{6} \left[ \frac{8N_{sockets}^{1.5} l^{2p-2}}{2p-2} - \frac{12N_{sockets} l^{2p-1}}{2p-1} + \frac{6N_{sockets}^{0.5} l^{2p}}{2p} - \frac{l^{2p+1}}{2p+1} \right]_{\sqrt{N_{sockets}}}^{2\sqrt{N_{sockets}}} \tag{A7}$$

And substituting the integration limits in (A7) we get:

$$p_2 = \frac{\alpha k \Gamma}{6} \left( \frac{2^{2p}-4}{p-1} + \frac{6-(1.5)2^{2p+1}}{p-0.5} + \frac{(3)2^{2p}-3}{p} + \frac{0.5-(0.5)2^{2p+1}}{p+0.5} \right) N_{sockets}^{p+0.5} \tag{A8}$$

Therefore,

$$p_1 + p_2 = \frac{\alpha k \Gamma}{6} \left( \left( \frac{1-(0.5)2^{2p+1}}{p+0.5} + \frac{(3)2^{2p}-6}{p} + \frac{9-(1.5)2^{2p+1}}{p-0.5} + \frac{2^{2p}-4}{p-1} \right) N_{sockets}^{p+0.5} \right.$$
$$\left. - \frac{0.5}{p+0.5} + \frac{3N_{sockets}^{0.5}}{p} - \frac{3N_{sockets}}{p-0.5} \right) \tag{A9}$$

Factorizing (A9), eliminating the constant $\frac{\alpha k \Gamma}{6}$ and rearranging will reduce to (A10):

$$p_1 + p_2 = N_{sockets}^{p+0.5} \frac{1.5(-p-1+2^{2p-1})}{(p+0.5)p(p-0.5)(p-1)} - \frac{0.5}{p+0.5} + \frac{3\sqrt{N_{sockets}}}{p} - \frac{3N_{sockets}}{p-0.5} \tag{A10}$$

Taking a common factor of $\frac{3\sqrt{N_{sockets}}}{(p-0.5)}$ from (A10) leads to (A11):

$$p_1 + p_2 = \frac{3\sqrt{N_{sockets}}}{(p-0.5)} \left( N_{sockets}^p \frac{0.5(-p-1+2^{2p-1})}{(p+0.5)p(p-1)} - \frac{(p-0.5)}{6\sqrt{N_{sockets}}(p+0.5)} + \frac{p-0.5}{p} - \sqrt{N_{sockets}} \right) \tag{A11}$$

(A11) is the numerator of (A2), and the denominator is shown below in (A12) that can be similarly evaluated as illustrated below:

$$\int_1^{2\sqrt{N_{sockets}}} i(l) dl = \frac{\alpha k \Gamma}{2} \int_1^{\sqrt{N_{sockets}}} \left( \frac{l^3}{3} - 2l^2 \sqrt{N_{sockets}} + 2lN_{sockets} \right) l^{2p-4} dl + \frac{\alpha k \Gamma}{6} \int_{\sqrt{N_{sockets}}}^{2\sqrt{N_{sockets}}} \left( 2\sqrt{N_{sockets}} - l \right)^3 l^{2p-4} dl \tag{A12}$$

Let the first and second sections of the above integral be $S_1$ and $S_2$ respectively, then we have:

$$S_1 = \frac{\alpha k \Gamma}{2} \left[ \frac{l^{2p}}{6p} - \frac{2\sqrt{N_{sockets}} l^{2p-1}}{2p-1} + \frac{2N_{sockets} l^{2p-2}}{2p-2} \right]_1^{\sqrt{N_{sockets}}} \tag{A13}$$

$$S_2 = \frac{\alpha k \Gamma}{6} \left[ \frac{8 N_{sockets}^{1.5} l^{2p-3}}{2p-3} - \frac{12 N_{sockets} l^{2p-2}}{2p-2} + \frac{6\sqrt{N_{sockets}} l^{2p-1}}{2p-1} - \frac{l^{2p}}{2p} \right]_{\sqrt{N_{sockets}}}^{2\sqrt{N_{sockets}}} \quad (A14)$$

Substituting the limits in (A13) and (A14) will give the results shown in (A15) and (A16) respectively:

$$S_1 = \frac{\alpha k \Gamma}{6} \left( \left( \frac{0.5}{p} - \frac{3}{p-0.5} + \frac{3}{p-1} \right) N_{sockets}^p - \frac{0.5}{p} + \frac{3 N_{sockets}^{0.5}}{p-0.5} - \frac{3 N_{sockets}}{p-1} \right) \quad (A15)$$

$$S_2 = \frac{\alpha k \Gamma}{6} \left( \frac{4(2^{2p-3}-1)}{p-1.5} + \frac{6(1-2^{2p-2})}{p-1} + \frac{3(2^{2p-1}-1)}{p-0.5} + \frac{0.5(1-2^{2p})}{p} \right) N_{sockets}^p \quad (A16)$$

Adding (A15) and (A16) gives:

$$S_1 + S_2 = \frac{\alpha k \Gamma}{6} \left( \left( \frac{1-(0.5)2^{2p}}{p} + \frac{(3)2^{2p-1}-6}{p-0.5} + \frac{9-(1.5)2^{2p}}{p-1} + \frac{2^{2p-1}-4}{p-1.5} \right) N_{sockets}^p \right.$$
$$\left. - \frac{0.5}{p} + \frac{3 N_{sockets}^{0.5}}{p-0.5} - \frac{3 N_{sockets}}{p-1} \right) \quad (A17)$$

We will now factorize (A17) after eliminating the constant $\frac{\alpha k \Gamma}{6}$ and rearranging gives (A18)

$$S_1 + S_2 = -N_{sockets}^p \frac{0.75(1+2p-2^{2p-1})}{p(p-0.5)(p-1)(p-1.5)} - \frac{0.5}{p} + \frac{3\sqrt{N_{sockets}}}{p-0.5} - \frac{3 N_{sockets}}{p-1} \quad (A18)$$

(A18) can be rewritten as shown in (A19):

$$S_1 + S_2 = 3 \left( -N_{sockets}^p \frac{(1+2p-2^{2p-1})}{p(2p-1)(p-1)(2p-3)} - \frac{1}{6p} + \frac{2\sqrt{N_{sockets}}}{2p-1} - \frac{N_{sockets}}{p-1} \right) \quad (A19)$$

Finally dividing (A11) by (A19) and rearranging gives (A20):

$$L_{avg} = \frac{\sqrt{N_{sockets}}}{(p-0.5)} \frac{\left( \frac{p-0.5}{p} - \sqrt{N_{sockets}} - \frac{(p-0.5)}{6\sqrt{N_{sockets}}(p+0.5)} + N_{sockets}^p \frac{(-p-1+2^{2p-1})}{2(p+0.5)p(p-1)} \right)}{\left( -N_{sockets}^p \frac{(1+2p-2^{2p-1})}{p(p-1)(2p-1)(2p-3)} - \frac{1}{6p} + \frac{2\sqrt{N_{sockets}}}{2p-1} - \frac{N_{sockets}}{p-1} \right)} \quad (A20)$$

Note the factor $\frac{\sqrt{N_{sockets}}}{(p-0.5)}$ which was not present in Sekars' derivation, and $L_{avg}$ here is in gate socket lengths.